\documentclass[english]{article}
\usepackage[T1]{fontenc}
\usepackage[latin9]{inputenc}
\usepackage{geometry}
\usepackage{xcolor}
\usepackage{array}
\usepackage{multirow}
\usepackage{graphicx}
\usepackage{float}
\usepackage[export]{adjustbox}[2011/08/13]
\makeatletter

\providecommand{\tabularnewline}{\\}
\newcommand{\lyxaddress}[1]{
	\par {\raggedright #1
	\vspace{1.4em}
	\noindent\par}
}

\makeatother

\usepackage{babel}
\begin{document}
\title{United Nation Security Council in Quantum World: Experimental Realization
of Quantum Anonymous Veto Protocols using IBM Quantum Computer}
\author{Satish Kumar, Anirban Pathak}
\maketitle

\lyxaddress{\begin{center}
Jaypee Institute of Information Technology, A 10, Sector 62, Noida,
UP-201309, India
\par\end{center}}
\begin{abstract}
United Nation (UN) security council has fifteen members, out of which five permanent members of the council can use their veto power against
any unfavorable decision taken by the council. In certain situation,
a member using right to veto may prefer to remain anonymous. This
need leads to the requirement of the protocols for anonymous veto
which can be viewed as a special type of voting. Recently, a few protocols
for quantum anonymous veto have been designed which clearly show quantum
advantages in ensuring anonymity of the veto. However, none of the
efficient protocols for quantum anonymous veto have yet been experimentally
realized. Here, we implement 2 of those protocols for quantum anonymous
veto using an IBM quantum computer named IBMQ Casablanca and different
quantum resources like Bell, GHZ and cluster states. In this set of
proof-of-principle experiments, it's observed that using the present
technology, a protocol for quantum anonymous veto can be realized
experimentally if the number of people who can veto remains small
as in the case of UN council. Further, it's observed that Bell state
based protocol implemented here performs better than the GHZ/cluster
state based implementation of the other protocol in an ideal scenario
as well as in presence of different types of noise (amplitude damping,
phase damping, depolarizing and bit-flip noise). In addition, it's
observed that based on diminishing impact on fidelity, different noise
models studied here can be ordered in ascending order as phase damping,
amplitude damping, depolarizing, bit-flip.
\end{abstract}

\section{Introduction\label{sec:Introduction}}

Quantum mechanics often exposes us to phenomena or rules that are
counter-intuitive to our classical minds. Interestingly, these counter-intuitive
phenomena or rules having no classical analogue lead to quantum advantages.
In fact, quantum advantages have already been reported in different
contexts, including but not restricted to quantum metrology \cite{arvidsson2020quantum,giovannetti2011advances},
quantum computing \cite{arute2019quantum,montanaro2016quantum}, and
quantum communication \cite{shenoy2017quantum}. As a result, it's
now widely believed that the civilization is ready for the second
quantum revolution \cite{dowling2003quantum}. However, all the facets
where we may be benefited from the uses of quantum resources are not
equally matured at the moment. To be precise, quantum cryptography
\cite{shenoy2017quantum} which can offer unconditional security,
seems to be most mature quantum technology at the moment as various
commercial solutions are already available. The primary feature of
quantum cryptography is quantum key distribution (QKD) which allows
generation and distribution of an unconditionally secure key between
two authentic parties using quantum resources. This field started
with the advent of the BB84 protocol \cite{BB_1984} in 1984. Since then considerable progress has happened. Specifically,
several schemes of QKD have been proposed and implemented \cite{ekert1991quantum}
and generalizing the basic idea of QKD, quantum advantage of obtaining
unconditional security has been extended to several new schemes for
one-way and two-way secure direct quantum communication (cf. Refs. \cite{pathak2015efficient,banerjee2012maximally,srikara2020continuous}
and Chapter 8 of \cite{pathak2013elements}) as well as schemes for
secure multiparty quantum computation (SMQC) \cite{saxena2020continuous,crepeau2002multiparty}.
Secure multiparty computation tasks include voting, auction, private
comparison, etc. As these tasks have enormous uses in our daily life,
protocols for SMQC have drawn considerable attention. Voting is one
of the most important SMQC tasks as it forms an integral part of any
democratic setup where decisions are taken collectively. The usefulness
of quantum resources for anonymous voting was first exploited in 2006
by Hillery et al. \cite{hillery2006towards} and almost simultaneously
by Vaccaro et al. \cite{vaccaro2007quantum} (for a historical note
see \cite{thapliyal2017protocols}). Since, then the field has grown
many a times with several new protocols appearing in the last few
years (\cite{thapliyal2017protocols,mishra2021quantumveto} and references
therein). 

Within the broad class of voting schemes, a specific subclass of interest
is one in which all the decisions are taken by unanimity only, i.e.,
a proposal put to the vote is rejected even if just one of the members
disagrees. One of the most prominent example of this is when a proposal
is put to vote in the United Nation (UN) security council, where the
proposal is rejected at once if one or more permanent member(s) of
the council disagrees. Such a scheme is known as the veto and many
times the situations require that the secrecy of the vote is maintained.
The first quantum solution for anonymous veto was provided by Rahman
and Kar by utilizing the GHZ correlations \cite{rahaman2015ghz}.
Following that Wang et al., provided a mature solution using the same
GHZ states and experimentally testing it in the case of four voters
on the IBM quantum computer \cite{wang2021anonymous}. Recently, Mishra
et al. \cite{mishra2021quantumveto} proposed a set of new schemes
for quantum anonymous veto and classified the schemes based on quantum
resources used along with their feasibility under realistic physical
implementations. Further, the schemes proposed by Mishra et al., have
been classified into probabilistic, iterative and deterministic schemes.
From the set of proposed schemes, two schemes (referred to as QAV-6
and QAV-7 in \cite{mishra2021quantumveto} and to be referred to as
Protocol A and Protocol B, respectively in this article) are of particular
interest as they have the highest efficiency. Here, we implement the
two most efficient quantum anonymous veto protocols proposed by Mishra
et al. \cite{mishra2021quantumveto} on the cloud-based IBM quantum
computer and investigate the effect of amplitude damping, phase damping,
bit flip and depolarizing noise on these protocols. 

Before proceeding further, it will be apt to note that due to the
presence of decoherence a scalable quantum computer cannot be build
until now. However, a set of noisy intermediate-scale quantum (NISQ)
computers have been built in the recent past. Further, cloud based
access of such quantum computers has also been provided by different
organizations. IBM is a pioneer in this aspect as they are providing
cloud-based access to NISQ computers to common researchers since 2016.
IBM's quantum computers are superconductivity based and utilizes transmon
qubits. However, the available computers have different topology and
number of qubits. Without going into those details, we may note that
IBM quantum computers have already been used for the realization of
various quantum computing and communication tasks \cite{saxena2021hybrid,sisodia2017experimental,sisodia2017design,harper2019fault,acasiete2020implementation}.
For example, IBM quantum computers are used in: quantum part of a
quantum-classical hybrid algorithm for factorization \cite{saxena2021hybrid},
nondestructive discrimination of Bell states \cite{sisodia2017experimental},
a proof-of-principle experiment for the implementation of an optimal
scheme for the teleportation of an $n$-qubit quantum state \cite{sisodia2017design},
the implementation of fault-tolerant logic gates in the code space
\cite{harper2019fault}, modeling quantum walk \cite{acasiete2020implementation}
and various other tasks (see \cite{behera2019designing} and references
therein).\textcolor{red}{{} }Extending this long list, in this paper,
we aim to report experimental realization of two efficient schemes
of QAV using IBM quantum computer. The motivation behind performing
such an experiment is two fold, firstly anonymous veto has many applications
in social life and secondly as veto usually involve a small number
of voters so commercial implementation QAV would be possible if it
can be successfully implemented using IBM quantum computer.

The rest of the paper is organized as follows. In Section \ref{sec:Quantum-veto-protocols},
we briefly describe the two most efficient quantum anonymous veto
protocols proposed by Mishra et al. \cite{mishra2021quantumveto}.
Then, we discuss the results and issues associated with the experimental
realization of those schemes on the IBM quantum computer in Section
\ref{sec:Experimetal-realization-of}. In Section \ref{sec:Comparison-of-Protocol},
performance of the experimentally realized protocols are compared
in the ideal scenario as well as in the presence of noise. Finally, the
paper is concluded in Section \ref{sec:Conclusion}. 

\section{Quantum anonymous veto protocols\label{sec:Quantum-veto-protocols}}

We have already mentioned that recently Mishra et al., \cite{mishra2021quantumveto}
have reported a set of schemes for QAV. Two of the reported schemes
are relatively more efficient (as far as qubit efficiency is concerned)
compared to the existing protocols and the other protocols proposed
by Mishra et al. In what follows, we briefly describe those two protocols
in a step-wise manner and refer to them as Protocols A and B. Both
the protocols are conducted by a semi-honest voting authority (VA)
referred to as Alice who is semi-honest in the sense that she strictly (i.e., honestly) follows the protocol, but tries to obtain additional information about the inputs (say, votes in our context) of the other users (i.e, voters). Voting involves $n$ voters $\left\{ V_{i},V_{2},\cdots,V_{n}\right\} .$

\subsection{Protocol A}

Protocol A is an iterative protocol and uses a Bell state for its
implementation with one qubit acting as a home qubit while the other
qubit acting as the travel qubit. The protocol can be described in
the following steps:
\begin{description}
\item [{\textcolor{black}{Step~A1}}] \textcolor{black}{VA} prepares a
maximally entangled Bell state ($\vert\phi^{+}\rangle=\frac{1}{\sqrt{2}}(|00\rangle+|11\rangle)$),
and keeps the first qubit (home qubit) with herself while sends the
second qubit (travel qubit) to the first voter ($V_{1})$.
\item [{\textcolor{black}{Step~A2}}] $V_{1}$ applies $\sigma_{z}(t)=\left[\begin{array}{cc}
1 & 0\\
0 & e^{i\frac{\pi}{2^{t}}}
\end{array}\right]$ with $t=0$ if he wishes to perform a veto, otherwise he applies
Identity operation $I=\left[\begin{array}{cc}
1 & 0\\
0 & 1
\end{array}\right].$ After the application of unitary on the travel qubit, $V_{1}$ sends
the travel qubit to $V_{2}$, who encodes his vote in the similar
manner and subsequently sends the travel qubit to $V_{3}$, and the
process continues till $V_{n}$ finally sends the travel qubit to
VA after executing his voting right.\\
Note that $(t+1)$ is the number of iterations
of the protocol. Thus, $t=0$ refers to the first iteration\footnote{Thus, in this iterative protocol to registrar a veto in $n$th
iteration a voter would apply $\sigma_{z}(n-1)=\left[\begin{array}{cc}
1 & 0\\
0 & e^{i\frac{\pi}{2^{n-1}}}
\end{array}\right]$.} and in the first iteration a voter would apply $\left[\begin{array}{cc}
1 & 0\\
0 & e^{i\pi}
\end{array}\right]=\left[\begin{array}{cc}
1 & 0\\
0 & -1
\end{array}\right]=\sigma_{z}$. 
\item [{Step~A3}] VA performs a Bell measurement using the home qubit
available with him and the travel qubit received from $V_{n}$. \\
If the veto is applied by an odd (even including 0) number of voters,
measurement of VA would yield $\vert\phi^{-}\rangle=\frac{1}{\sqrt{2}}(|00\rangle-|11\rangle)$
$\left(\vert\phi^{+}\rangle=\frac{1}{\sqrt{2}}(|00\rangle+|11\rangle)\right)$
and the protocol will be accomplished (continued to the next step as the
result is inconclusive). 
\item [{Step~A4}] Steps A1-A3 are repeated for $t=1$ and so on till one
gets a conclusive result with each iteration increasing the value
of $t$ by one.
\end{description}
If $n$ number of voters participate in the process, then the maximum
number of iterations required to arrive on conclusive result would
be $1+\log_{2}n$ with every iteration eliminating half of the voting
possibilities.

\subsection{Protocol B}

In contrast to the previous protocol, this protocol is a deterministic
protocol, where the conclusive result is obtained in a single iteration
only. Further, every voter is provided with a set of unitaries, which
can be used to give a veto. Specifically, $i$th user possesses a
unitary $U_{i}$ which he can apply to register a veto. As a specific
example, possible unitary operations for 4 voters and different entangled
states is shown in Table \ref{tab:table 1}. The steps involved in
executing this protocol for $n$ number of voters in general are as
follows: 
\begin{description}
\item [{Step~B1}] \textcolor{black}{VA} prepares a $m$-qubit entangled
state $\mid\psi_{in}\rangle$ (with $m\geq(n-1)c$). \\
Here, $c=1$ as voters can encode only 1 bit of information either
in favor or against the proposal. \textcolor{black}{VA} further makes
$l$ qubits ($l<m$) as travel qubits and keeps $\left(m-l\right)$
qubits as the home qubits with $l$ qubits traveling to each of the
voter and finally returned back to VA.
\item [{Step~B2}] The voter $V_{i}$ $(1\leq i\leq n-1)$ applies the
identity operation when he is in favor and apply an operator $U_{i}$
when he is in against. After the operation, the $l$ travel qubits
will be sent to voter $V_{i+1}$. This process continues till the
voter $V_{n}$ who sends the travel qubits to VA after his operation. 
\item [{Step~B3}] \textcolor{black}{VA} measures the final state (say
$|\psi_{fin}\rangle$ ) in the same basis in which initial state was
prepared. If $\langle\psi_{in}|\psi_{fin}\rangle=1$ then it is concluded
that either no voter has done veto or all voters have done veto.
If $\langle\psi_{in}|\psi_{fin}\rangle\neq1$ then it is concluded
that at least one voter has applied the veto. In this way, the voters
can conclude if there is any consensus or not. 
\begin{table}
\begin{centering}
\begin{tabular}{|>{\centering}p{2cm}|>{\centering}p{2cm}|>{\centering}p{2cm}|>{\centering}p{9cm}|}
\hline 
Number of voters & Number of travel qubits & Quantum state & Encoding operation ($O_{i}$)\tabularnewline
\hline 
4 & 2 & 4-qubit cluster state & $U_{1}:\{X\otimes iY\},U_{2}:\{X\otimes Z\},U_{3}:\{iY\otimes Z\},U_{4}:\{iY\otimes iY\}$\tabularnewline
\hline 
4 & 2 & GHZ state & $U_{1}:\{X\otimes I\},U_{2}:\{X\otimes X\},U_{3}:\{iY\otimes X\},U_{4}:\{iY\otimes I\}$\tabularnewline
\hline 
\end{tabular} 
\par\end{centering}
\centering{}\caption{A few examples of the quantum states and the corresponding quantum operations
that may be used to implement Protocol B.\label{tab:table 1}}
\end{table}
\end{description}

\section{Quantum veto protocols using IBM quantum computer\label{sec:Experimetal-realization-of}}

\begin{figure}
\begin{centering}
\includegraphics{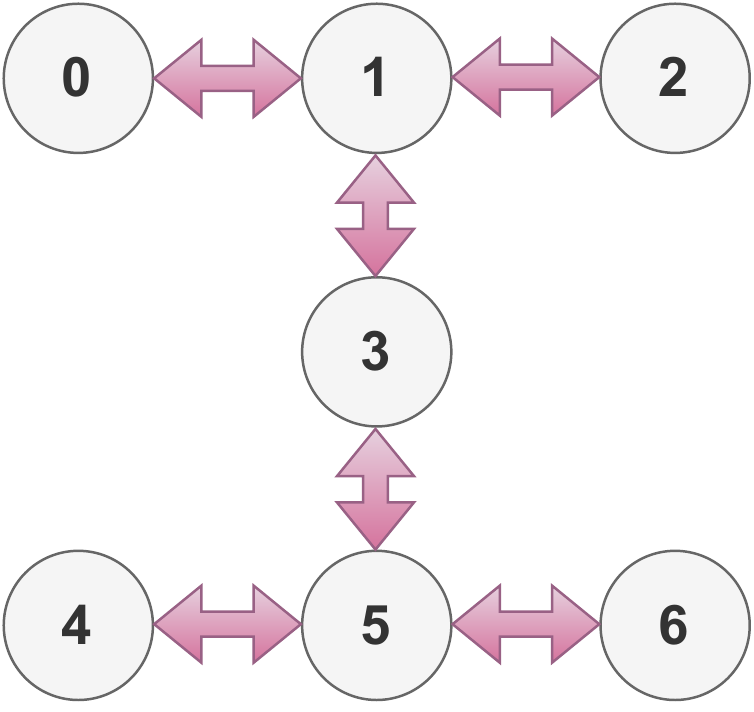}
\par\end{centering}
\caption{(Color online)Topology of the IBMQ Casablanca\label{fig:(Color-online)Topology-ofIBM}.}
\end{figure}

\textcolor{black}{As mentioned above IBM provides cloud based access
to a set of quantum computers, which are distinct from each other
as far as their size (number of qubits) and topology are concerned
\cite{IBMQ(2021)}. Without loss of generality, for the experimental
realization of the above described protocols, here we have considered
the number of voters as four. Further, Protocol A requires Bell states
only so any small quantum computer can in principle be used for proof-of-principle
implementation of the Protocol A. In contrast, Protocol B can be implemented
using a class of entangled states. Of course the choice of unitary
operations would depend on the choice of the entangled state. In Table
\ref{tab:table 1}, we have already described the unitaries to be
used by the voters if a 4 qubit cluster state or a 3 qubit GHZ state
is to be used for the implementation of Protocol B. Here, we will restrict
us to realize Protocol B using these entangled states only. Thus,
any quantum computer which can perform quantum information processing
involving four or more qubits will be sufficient for our purpose.
Here, we have chosen }IBMQ Casablanca, because of its availability
at the time of performing these experiments. It may be noted that
IBMQ Casablanca is a 7 qubit superconductivity-based quantum computer
whose topology is shown in Fig. \ref{fig:(Color-online)Topology-ofIBM}.
In what follows, we briefly describe the quantum circuits used for
the realization of the protocols of quantum veto and the results obtained
by implementing those circuits in IBMQ Casablanca. A two qubit gate
(say a CNOT gate) can be directly implemented only between the qubits
which are connected through an arrow. A bidirectional arrow implies
that any one of the qubits shown at the end of the arrow can work
as control qubit. However, every CNOT gate does not work with the
same accuracy. Errors introduced by a CNOT gate depends on the choice
of the pair of qubits on which the CNOT gate is
applied. To be explicit on this point, in Table \ref{tab:Description-of-errors}
we list the experimental parameters and errors in different gate implementation
in IBMQ Casablanca that was present at the time of performing this
experiment. We restrict the table to qubit 0,1,2,3 shown in Fig. \ref{fig:(Color-online)Topology-ofIBM}
and marked as $Q_{0},Q_{1},Q_{2},Q_{3}$, respectively in Table \ref{tab:Description-of-errors}
as we have only used these qubits for our experiment. In fact, to
implement Protocol A we have used $Q_{0}$ and $Q_{1}$ and to implement
Protocol B using GHZ states we have used qubits $Q_{0}, Q_{1}$ and $Q_{2}$, whereas to implement Protocol B using 4-qubit cluster state we have used all
the four qubits $\{Q_{0},Q_{1},Q_{2},Q_{3}\}$.

\begin{table}
\begin{centering}
\begin{tabular}{|c|c|c|>{\centering}p{2cm}|>{\centering}p{2cm}|>{\centering}p{2cm}|>{\centering}p{5cm}|}
\hline 
Qubit & T1 ($\mu s$) & T2 ($\mu s$) & \centering{}Frequency (GHz) & \centering{}Readout assignment error & \centering{}Single-qubit Pauli-X-error & \centering{}CNOT error\tabularnewline
\hline 
$Q_{0}$ & 108.61 & 38.65 & \centering{}4.822 & \centering{}$3.74\times10^{-2}$ & \centering{}$2.531\times10^{-4}$ & \centering{}cx0\_1: $1.081\times10^{-2}$\tabularnewline
\hline 
$Q_{1}$ & 113.03 & 70.78 & \centering{}4.76 & \centering{}$2.68\times10^{-2}$ & \centering{}$2.012\times10^{-4}$ & \centering{}cx1\_3: $6.945\times10^{-3}$, cx1\_2: $9.599\times10^{-3}$,
cx1\_0: $1.081\times10^{-2}$\tabularnewline
\hline 
$Q_{2}$ & 95.43 & 125 & \centering{}4.906 & \centering{}$2.05\times10^{-2}$ & \centering{}$2.716\times10^{-4}$ & \centering{}cx2\_1: $9.599\times10^{-3}$\tabularnewline
\hline 
$Q_{3}$ & 102.59 & 126.79 & \centering{}4.879 & \centering{}$2.99\times10^{-2}$ & \centering{}$4.012\times10^{-4}$ & \centering{}cx3\_1: $6.945\times10^{-3}$\tabularnewline
\hline 
\end{tabular}
\par\end{centering}
\caption{Description of the relevant errors in the quantum processor used (i.e.,
IBMQ Casablanca). The table is prepared using the calibration data
available at the website of IBM quantum services at the time of performing
the experiments (on October 13, 2021). Here, cxi\_j represents the error
in the implementation of a CNOT gate, where index i denotes a control
qubit and the index j denotes a target qubit.\label{tab:Description-of-errors}}
\end{table}

\subsection{Experimental realization of Protocol A}

\begin{figure}
\begin{centering}
\includegraphics[scale=0.9]{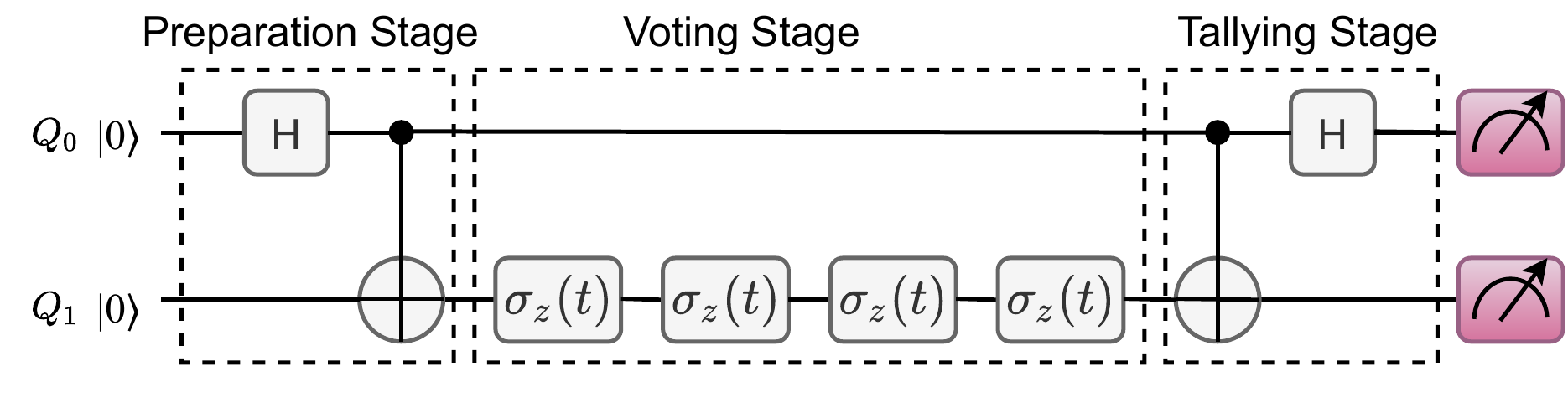}
\par\end{centering}
\centering{}\caption{(Color online) A quantum circuit for the experimental realization
of Protocol A. \label{fig:A-QV_bell}}
\end{figure}

\textcolor{black}{To implement Protocol A, VA needs to prepare a Bell
state $|\phi_{int}\rangle=\frac{1}{\sqrt{2}}(|00\rangle+|11\rangle)$
which can be prepared using a CNOT gate followed by a Hadamard
gate (cf. leftmost block of Fig \ref{fig:A-QV_bell}). Subsequently,
voters as well as VA follow the protocol and to do so 4 voters need
to apply unitaries in sequence, where $i$th voter $V_{i}$ applies
$\sigma_{z}(t)$ and specific choice of $\sigma_{z}(t)$ depends on
the number of iterations and whether a voter is executing his right
to veto. A general circuit representation of this part is shown in
the middle part of Fig \ref{fig:A-QV_bell}. This step would produce
a final state $|\phi_{fin}\rangle$ which is produced after all voters
have applied their votes in a particular iteration round. If $\langle\phi_{int}|\phi_{fin}\rangle=1$
then the VA remains inconclusive, otherwise she can conclude that
everyone does not agree with the proposal and veto has been applied.
Now, to check whether the final state satisfies $\langle\phi_{int}|\phi_{fin}\rangle=1$
or not, one needs to measure $|\phi_{fin}\rangle$ in Bell basis,
but IBM quantum computers do not allow direct measurement in any basis
other than computational basis. Consequently a reverse EPR circuit
comprising of a Hadamard gate followed by the CNOT gate is used for transforming
Bell measurement into computational basis measurement (see rightmost
block of Fig \ref{fig:A-QV_bell}). Thus, Fig. \ref{fig:A-QV_bell}
describes the general structure of a complete circuit that can be
used to implement the quantum veto protocol in }IBMQ Casablanca.\textcolor{black}{{}
Since, we have considered four voters, the maximum number of iterations
of the protocol to get a conclusive result would be $(1+\log_{2}4=3)$. }

One can easily see that here we have following five possible voting
patterns: no one vetoed, any one voter out of the total 4 voters vetoed,
any two of the 4 voters vetoed, any 3 of the 4 voters vetoed, and
all the voters have vetoed. Expected outcomes of all these possibilities
after each iteration are listed in the 6th column of the Table \ref{tab:Result Bell}.
The task remained is to check whether a real experimental run using
the circuit shown in Fig. \ref{fig:A-QV_bell} in IBMQ Casablanca
yields results consistent with the expected results listed in the
6th column as the final state of Table \ref{tab:Result Bell} (or
in the 8th column of the Table \ref{tab:Result Bell} as the corresponding
expected outcome of measurement when the measurement in Bell basis
is transformed to measurement in computational basis).

\begin{table}
\centering{}%
\begin{tabular}{|>{\centering}p{0.5cm}|>{\centering}p{1cm}|>{\centering}p{1.1cm}|>{\centering}p{2.2cm}|>{\centering}p{1.5cm}|>{\centering}p{1cm}|>{\centering}p{1.64cm}|>{\centering}p{1.5cm}|>{\centering}p{2cm}|>{\centering}p{1cm}|}
\hline 
Case & Initial state & Number of veto & Which voter(s) has (have) vetoed & Iteration No. & Final state & Result & Simulator result (or expected measurement outcome)  & Probability of obtaining the expected result on real device & Fidelity (\%)\tabularnewline
\hline 
1 & $|\phi^{+}\rangle$ & 0 & No one & Iteration 1 & $|\phi^{+}\rangle$ & Inconclusive & 00 & 0.985 & 99.41\tabularnewline
\hline 
2 & $|\phi^{+}\rangle$ & 1 & Any one voter among the 4 voters & Iteration 1 & $|\phi^{-}\rangle$ & Conclusive & 10 & 0.878 & 96.50\tabularnewline
\hline 
3 & $|\phi^{+}\rangle$ & 2 & \multirow{2}{2.5cm}{Any two of the 4 voters (e.g., $1^{st}$ \& $3^{rd}$ or $3^{rd}$
\& $4^{th}$)} & Iteration 1 & $|\phi^{+}\rangle$ & Inconclusive & 00 & 0.981 & 98.65\tabularnewline
\cline{5-10} \cline{6-10} \cline{7-10} \cline{8-10} \cline{9-10} \cline{10-10} 
 &  &  &  & Iteration 2 & $|\phi^{-}\rangle$ & Conclusive & 10 & 0.911 & 96.23\tabularnewline
\hline 
4 & $|\phi^{+}\rangle$ & 3 & Any three of the 4 voters (e.g., $1^{st}$, $2^{nd}$ \& $4^{th}$
or$1^{st}$, $3^{rd}$ \& $4^{th}$) & Iteration 1 & $|\phi^{-}\rangle$ & Conclusive & 10 & 0.915 & 98.44\tabularnewline
\hline 
5 & $|\phi^{+}\rangle$ & 4 & All the four voters & Iteration 1 & $|\phi^{+}\rangle$ & Inconclusive & 00 & 0.980 & 98.98\tabularnewline
\cline{5-10} \cline{6-10} \cline{7-10} \cline{8-10} \cline{9-10} \cline{10-10} 
 &  &  &  & Iteration 2 & $|\phi^{+}\rangle$ & Inconclusive & 00 & 0.981 & 99.44\tabularnewline
\cline{5-10} \cline{6-10} \cline{7-10} \cline{8-10} \cline{9-10} \cline{10-10} 
 &  &  &  & Iteration 3 & $|\phi^{-}\rangle$ & Conclusive & 10 & 0.964 & 95.19\tabularnewline
\hline 
\end{tabular}\caption{Comparison of the theoretically expected results for the realization
of quantum veto using Bell state (Protocol A) with the corresponding
results obtained experimentally. Here, $|\phi^{\pm}\rangle=\frac{1}{\sqrt{2}}\left(|00\rangle+|11\rangle\right)$
and simulator result corresponds to the measurement outcome obtained
on execution of the circuit shown in Fig \ref{fig:A-QV_bell} in IBM
Qasm. \label{tab:Result Bell}}
\end{table}

\begin{figure}
\begin{centering}
\includegraphics[scale=0.5]{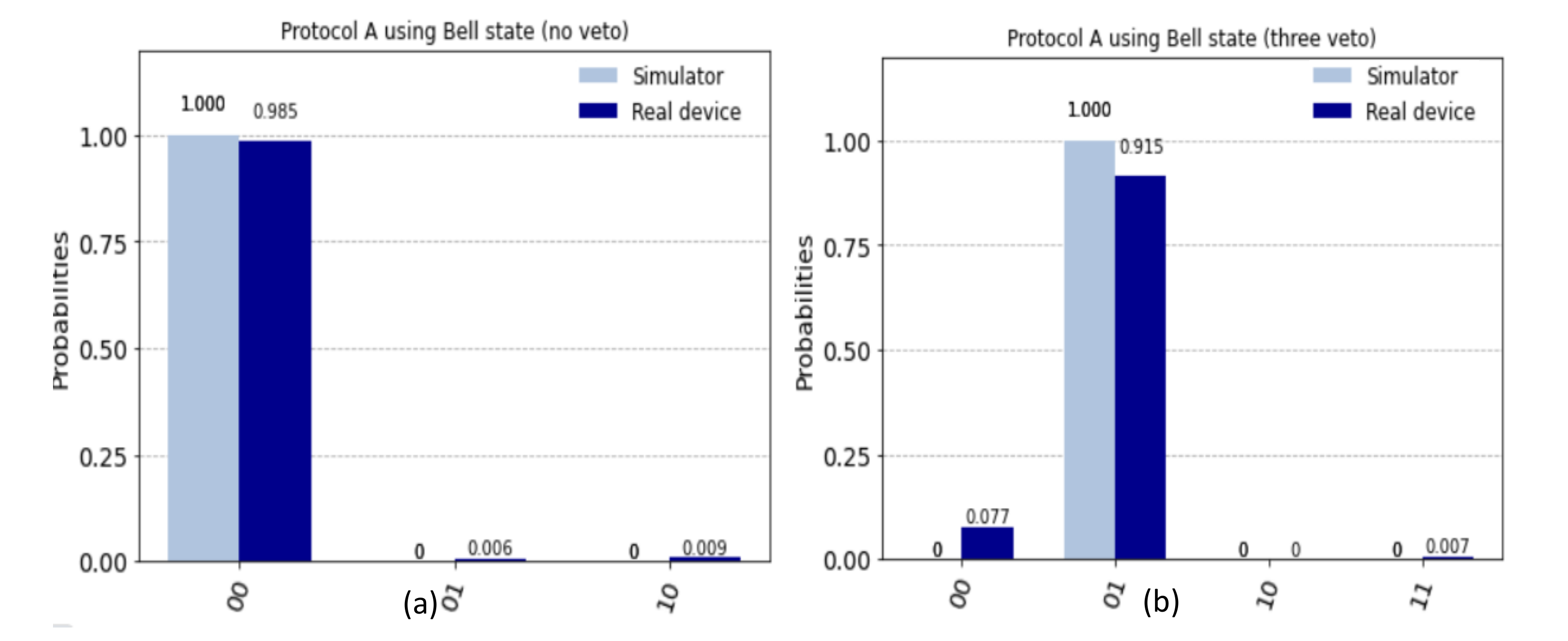}
\par\end{centering}
\caption{\label{fig:QV bell} (Color online) Simulator and real device results
for Protocol A using Bell state for the (a) inconclusive and (b) conclusive
results.\textcolor{blue}{{} \label{fig:-(Protocol A result}}}
\end{figure}

We have experimentally executed the circuit shown in Fig. \ref{fig:A-QV_bell}
in IBMQ Casablanca. In addition, we executed the circuit in the IBM
Qasm simulator, too. The number of shots chosen to run each experiment
in real device is 8192. As expected the simulation results are found
to be in perfect consistency with the expected theoretical results
listed in the 6th column of the Table \ref{tab:Result Bell}.
However, due to experimental limitations reflected in the errors in
implementing different gates and in the readout of the final results
(cf. Table \ref{tab:Description-of-errors}), the experimental results are
found to be slightly different from the expected theoretical results
(see last two columns of Table \ref{tab:Result Bell}). Specifically,
in the last to last column of Table \ref{tab:Result Bell}, we only
report the probability of obtaining the desired result. In real experiment,
a different state is produced and its closeness with the expected
quantum state is quantified through fidelity which is defined as
$F(\sigma,\rho)=Tr\left[\sqrt{{\sqrt{{\sigma}}\rho\sqrt{{\sigma}}}}\right]^{2}$, where $\sigma$ is theoretical final state density matrix and $\rho$
is experimental final state density matrix. The fidelity is obtained
using an inbuilt feature of IBM quantum experience which actually
performs quantum state tomography to obtain the density matrix
of the output state and subsequently uses the above mentioned expression
for fidelity to compute it. As it can be seen in the last column of
Table \ref{tab:Result Bell}, fidelity ranges between\textcolor{purple}{{}
}95.19\% to 99.41\%, we may conclude that experimental outcomes are
pretty close to the expected outcome and with high success probability
quantum veto schemes can be implemented. To illustrate this point in
Fig. \ref{fig:-(Protocol A result} we show the output of experimental
realization of Protocol A for two specific cases along with the corresponding
simulation (or equivalently theoretical results). The closeness observed
in results is indicative of successful implementation of quantum anonymous
veto. 

In addition to the above, consistency of the obtained result is verified
by performing 10 independent runs (each having 8192 shots) of the
experiment for the situation where 3 of the 4 voters applied veto.
Fidelity (in \%) for these 10 independent runs of the experiment are
obtained as 98.44, 97.56, 96.19, 97.70, 97.32, 97.26, 95.03, 95.84,
96.33, 97.53. This dataset has a standard deviation of $\sim1.03$
which indicates that the results are quite consistent.

Here it may be noted that with the increase in the number of voters,
the size of the quantum circuit (number of gates in the circuit) will
increase as each voter will apply his unitary. As a consequence of
increased gate count, errors will increase because each gate will
introduce some error, and naturally fidelity will drop. However, the
quantum anonymous veto can be used in real life scenario (like UN
security council where only 5 voters have right to give veto) with
the existing technology as usually a small number of voters are involved
in a process of veto. Further, in a large set of voters, it's not
expected that all will agree on any proposal. 

\subsection{Experimental realization of Protocol B}

\begin{figure}
\begin{centering}
\includegraphics[scale=0.6]{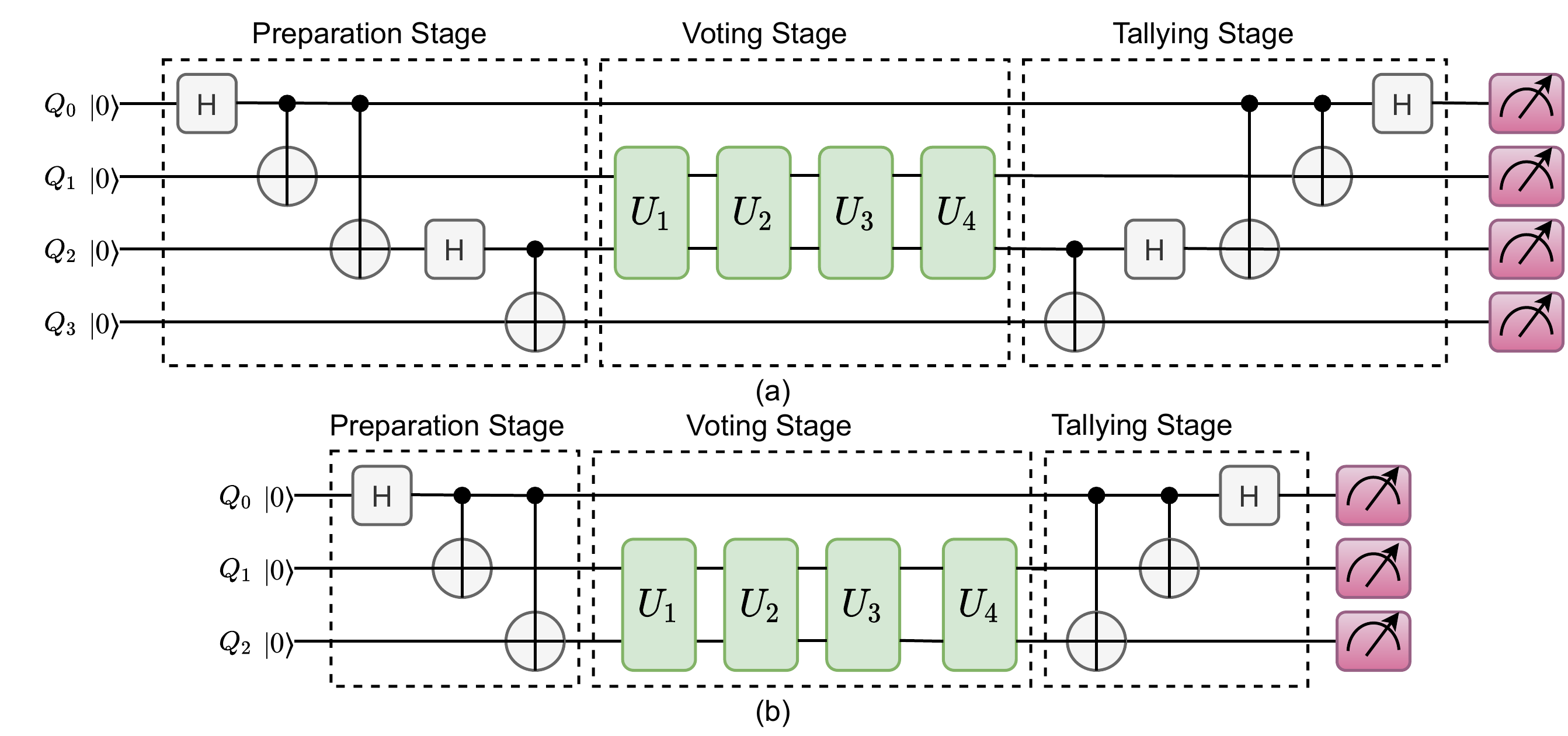}
\par\end{centering}
\caption{(Color online) A quantum circuit for realization of Protocol B using
the (a) cluster state and the (b) GHZ state. \label{fig:A-quantum-circuit}}
 
\end{figure}

\begin{figure}
\begin{centering}
\includegraphics[scale=0.6]{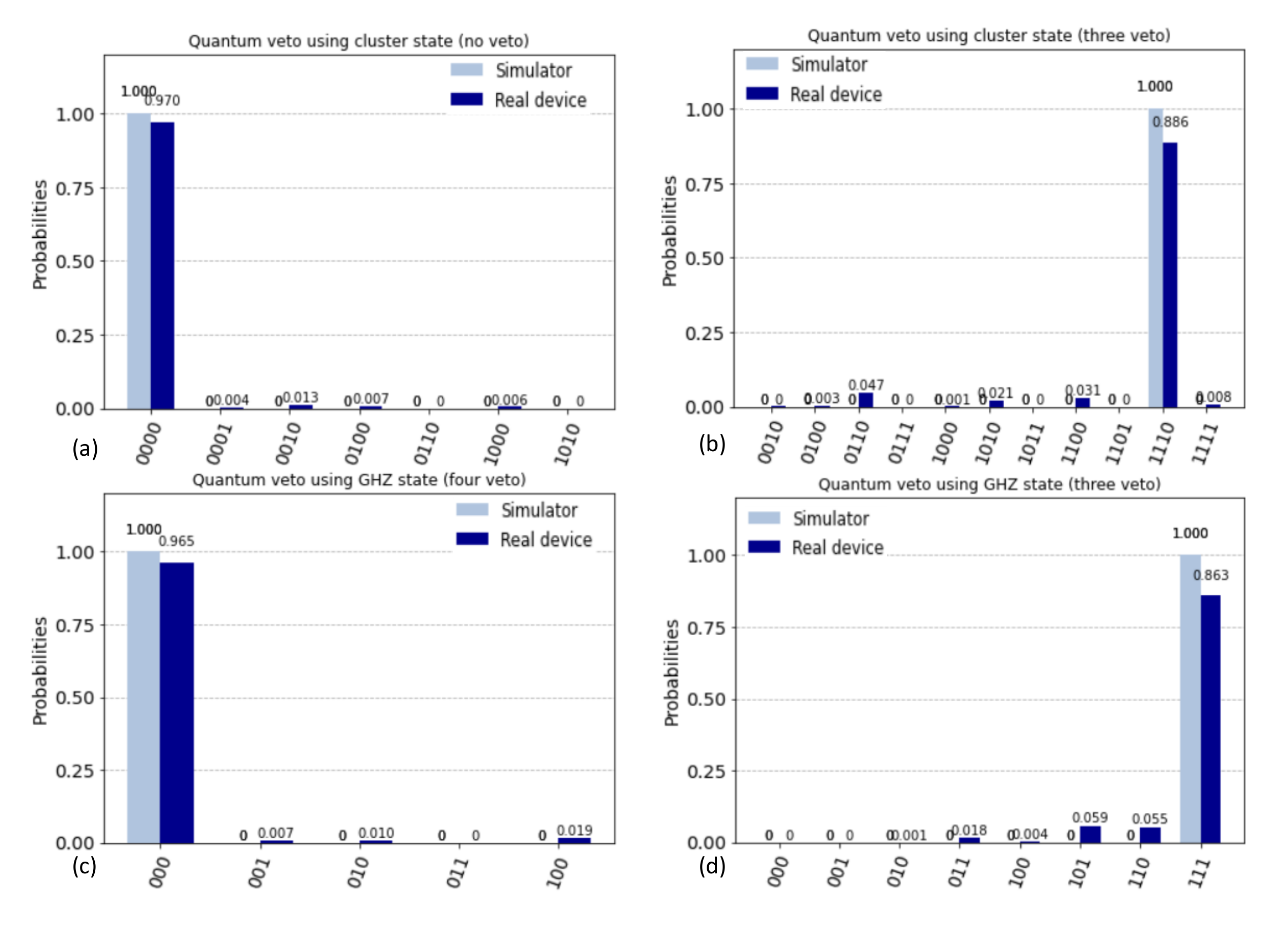}
\par\end{centering}
\caption{(Color online) Simulator and real device results for Protocol B using
cluster state and GHZ state for the (a,c) inconclusive and (b,d) conclusive
result. \label{fig:Protocol B}}
\end{figure}

Similar to Protocol A, we consider that there are 4 voters and the
circuits to be implemented in IBMQ Casablanca to have three parts.
Leftmost part to be used for preparation of the entangled state (see
the leftmost blocks of \ref{fig:A-quantum-circuit} (a) and (b), which
are used to produce 4-qubit cluster state $|\psi_{c}\rangle=\frac{1}{\sqrt{2}}(|0000\rangle+|0011\rangle+|1100\rangle-|1111\rangle)$
and 3-qubit GHZ state $|\psi_{GHZ}\rangle=\frac{1}{\sqrt{2}}(|000\rangle+|111\rangle)$,
respectively). The middle block is to be used for voting in accordance
with the rules described in Table \ref{tab:table 1}.\textcolor{black}{{}
The protocol is followed by all voters and finally the travel qubits
are returned back to VA. Now, in the right most block VA needs to
perform measurement in the same basis in which the initial entangled
state was prepared. However, such a measurement is not directly allowed
in IBM quantum computers, so we need a circuit which may be viewed
as the inverse of the circuit used in the preparation stage to transform
the measurement into a computational basis measurement. The same is
done in the rightmost block }of Fig \ref{fig:A-quantum-circuit} (a) and
(b)\textcolor{black}{, for the cluster state based implementation and
GHZ state based implementation, respectively. Expected results for
5 different possibilities of voting pattern using cluster (GHZ) states
are summarized in the first seven columns of Table \ref{tab:Result cluster}
(Table \ref{tab:Result ghz}). As expected these results of ideal
scenario are consistent with the IBM Qasm results. However, the actual
experimental realization led to results which are slightly different
from the ideal results and those are listed in the last two columns
of Table \ref{tab:Result cluster} and Table \ref{tab:Result ghz}
and also illustrated in Fig. \ref{fig:Protocol B}. }

Results depicted in Fig. \ref{fig:Protocol B} (a, b) (Fig. \ref{fig:Protocol B}
(c, d) for cluster state (GHZ state) based implementation of the Protocol
B using the IBMQ Casablanca along with the expected results of ideal scenario
clearly show that proof-of-principle realization of quantum anonymous
veto protocol (Protocol B) has happened successfully in IBMQ Casablanca
for 8192 repetitions (shots) of the experiment. 

\begin{table}
\begin{centering}
\footnotesize
\begin{tabular}{|>{\centering}p{0.5cm}|>{\centering}p{1cm}|>{\centering}p{1cm}|>{\centering}p{1.5cm}|c|>{\centering}p{1.8cm}|>{\centering}p{1cm}|>{\centering}p{1.5cm}|>{\centering}p{1.4cm}|}
\hline 
Case & Initial State & Number of veto & Which voter(s) has (have) vetoed & Final State & Is the result conclusive? & Simulator result (or expected measurement outcome) & Probability of getting expected result on real device & Fidelity (\%)\tabularnewline
\hline 
1 & $|\psi_{c}\rangle$ & 0 & No one & $\frac{1}{2}(|0000\rangle+|0011\rangle+|1100\rangle-|1111\rangle)$ & No & 0000 & 0.970 & 97.63\tabularnewline
\hline 
\multirow{4}{0.5cm}{\centering{2}} & \multirow{4}{1cm}{\centering{$|\psi_{c}\rangle$}} & \multirow{4}{1cm}{\centering{1}} & $1^{st}$ & $\frac{1}{2}(|0101\rangle-|0110\rangle-|1001\rangle-|1010\rangle)$ & Yes & 1111 & 0.873 & 89.06\tabularnewline
\cline{4-9} \cline{5-9} \cline{6-9} \cline{7-9} \cline{8-9} \cline{9-9} 
 &  &  & $2^{nd}$ & $\frac{1}{2}(|0100\rangle-|0111\rangle+|1000\rangle+|1011\rangle)$ & Yes & 0110 & 0.887 & 93.23\tabularnewline
\cline{4-9} \cline{5-9} \cline{6-9} \cline{7-9} \cline{8-9} \cline{9-9} 
 &  &  & $3^{rd}$ & $\frac{1}{2}(-|0100\rangle+|0111\rangle+|1000\rangle+|1011\rangle)$ & Yes & 1110 & 0.779 & 91.17\tabularnewline
\cline{4-9} \cline{5-9} \cline{6-9} \cline{7-9} \cline{8-9} \cline{9-9} 
 &  &  & $4^{th}$ & $\frac{1}{2}(-|0101\rangle+|0110\rangle-|1001\rangle-|1010\rangle)$ & Yes & 0111 & 0.819 & 88.79\tabularnewline
\hline 
\multirow{6}{0.5cm}{\centering{3}} & \multirow{6}{1cm}{\centering{$|\psi_{c}\rangle$}} & \multirow{6}{1cm}{\centering{2}} & $1^{st}$ \& $2^{nd}$ & $\frac{1}{2}(-|0001\rangle-|0010\rangle+|1101\rangle-|1110\rangle)$ & Yes & 1001 & 0.905 & 92.69\tabularnewline
\cline{4-9} \cline{5-9} \cline{6-9} \cline{7-9} \cline{8-9} \cline{9-9} 
 &  &  & $1^{st}$ \& $3^{rd}$ & $\frac{1}{2}(|0001\rangle+|0010\rangle+|1101\rangle-|1110\rangle)$ & Yes & 0001 & 0.941 & 95.32\tabularnewline
\cline{4-9} \cline{5-9} \cline{6-9} \cline{7-9} \cline{8-9} \cline{9-9} 
 &  &  & $1^{st}$ \& $4^{th}$ & $\frac{1}{2}(|0000\rangle+|0011\rangle-|1100\rangle+|1111\rangle)$ & Yes & 1000 & 0.926 & 94.95\tabularnewline
\cline{4-9} \cline{5-9} \cline{6-9} \cline{7-9} \cline{8-9} \cline{9-9} 
 &  &  & $2^{nd}$ \& $3^{rd}$ & $\frac{1}{2}(-|0000\rangle-|0011\rangle+|1100\rangle-|1111\rangle)$ & Yes & 1000 & 0.930 & 94.58\tabularnewline
\cline{4-9} \cline{5-9} \cline{6-9} \cline{7-9} \cline{8-9} \cline{9-9} 
 &  &  & $2^{nd}$ \& $4^{th}$ & $\frac{1}{2}(-|0001\rangle-|0010\rangle-|1101\rangle+|1110\rangle)$ & Yes & 0001 & 0.919 & 94.04\tabularnewline
\cline{4-9} \cline{5-9} \cline{6-9} \cline{7-9} \cline{8-9} \cline{9-9} 
 &  &  & $3^{rd}$ \& $4^{th}$ & $\frac{1}{2}(-|0001\rangle-|0010\rangle+|1101\rangle-|1110\rangle)$ & Yes & 1001 & 0.917 & 93.12\tabularnewline
\hline 
\multirow{4}{0.5cm}{\centering{4}} & \multirow{4}{1cm}{\centering{$|\psi_{c}\rangle$}} & \multirow{4}{1cm}{\centering{3}} & $1^{st}$, $2^{nd}$ \& $3^{rd}$ & $\frac{1}{2}(-|0101\rangle+|0110\rangle-|1001\rangle-|1010\rangle)$ & Yes & 0111 & 0.886 & 92.22\tabularnewline
\cline{4-9} \cline{5-9} \cline{6-9} \cline{7-9} \cline{8-9} \cline{9-9} 
 &  &  & $1^{st}$, $2^{nd}$ \& $4^{th}$ & $\frac{1}{2}(-|0100\rangle+|0111\rangle+|1000\rangle+|1011\rangle)$ & Yes & 1110 & 0.882 & 91.15\tabularnewline
\cline{4-9} \cline{5-9} \cline{6-9} \cline{7-9} \cline{8-9} \cline{9-9} 
 &  &  & $1^{st}$, $3^{rd}$ \& $4^{th}$ & $\frac{1}{2}(-|0100\rangle+|0111\rangle-|1000\rangle-|1011\rangle)$ & Yes & 0110 & 0.905 & 91.28\tabularnewline
\cline{4-9} \cline{5-9} \cline{6-9} \cline{7-9} \cline{8-9} \cline{9-9} 
 &  &  & $2^{nd}$, $3^{rd}$ \& $4^{th}$ & $\frac{1}{2}(-|0101\rangle+|0110\rangle+|1001\rangle+|1010\rangle)$ & Yes & 1111 & 0.829 & 88.89\tabularnewline
\hline 
5 & $|\psi_{c}\rangle$ & 4 & $1^{st}$, $2^{nd}$, $3^{rd}$ \& $4^{th}$ & $\frac{1}{2}(|0000\rangle+|0011\rangle+|1100\rangle-|1111\rangle)$ & No & 0000 & 0.964 & 96.85\tabularnewline
\hline 
\end{tabular}
\par\end{centering}
\caption{Results for the realization of quantum veto using the cluster state (Protocol B).
Here, $|\psi_{c}\rangle=\frac{1}{2}(|0000\rangle+|0011\rangle+|1100\rangle-|1111\rangle)$
\label{tab:Result cluster}}
\end{table}

\begin{table}
\begin{centering}
\footnotesize
\begin{tabular}{|>{\centering}p{1cm}|>{\centering}p{1.2cm}|>{\centering}p{1cm}|>{\centering}p{2cm}|>{\centering}p{3cm}|>{\centering}p{1.5cm}|>{\centering}p{1.6cm}|>{\centering}p{2cm}|>{\centering}p{1.5cm}|}
\hline 
Case & Initial State & Number of veto & Which voter(s) has (have) vetoed & Final State & Is the result conclusive? & Simulator result (or expected measurement outcome) & Probability of getting expected result on real device & Fidelity (\%)\tabularnewline
\hline 
\centering{1} & $\centering{|\psi_{GHZ}\rangle}$ & \centering{0} & No one & $\frac{1}{\sqrt{2}}(|000\rangle+|111\rangle)$ & No & 000 & 0.972 & 97.67\tabularnewline
\hline 
\centering{2} & $\centering{|\psi_{GHZ}\rangle}$ & \centering{1} & $1^{st}$  & $\frac{1}{\sqrt{2}}(|010\rangle+|101\rangle)$ & Yes & 010 & 0.916 & 94.79\tabularnewline
\cline{4-9} \cline{5-9} \cline{6-9} \cline{7-9} \cline{8-9} \cline{9-9} 
 &  &  & $2^{nd}$ & $\frac{1}{\sqrt{2}}(|011\rangle+|100\rangle)$ & Yes & 011 & 0.934 & 91.96\tabularnewline
\cline{4-9} \cline{5-9} \cline{6-9} \cline{7-9} \cline{8-9} \cline{9-9} 
 &  &  & $3^{rd}$ & $\frac{1}{\sqrt{2}}(-|011\rangle+|100\rangle)$ & Yes & 111 & 0.850 & 90.72\tabularnewline
\cline{4-9} \cline{5-9} \cline{6-9} \cline{7-9} \cline{8-9} \cline{9-9} 
 &  &  & $4^{th}$ & $\frac{1}{\sqrt{2}}(-|010\rangle+|101\rangle)$ & Yes & 110 & 0.900 & 93.44\tabularnewline
\hline 
\multirow{6}{1cm}{\centering{3}} & $|\psi_{GHZ}\rangle$ & 2 & $1^{st}$ \& $2^{nd}$ & $\frac{1}{\sqrt{2}}(|001\rangle+|110\rangle)$ & Yes & 001 & 0.912 & 94.52\tabularnewline
\cline{4-9} \cline{5-9} \cline{6-9} \cline{7-9} \cline{8-9} \cline{9-9} 
 &  &  & $1^{st}$ \& $3^{rd}$ & $\frac{1}{\sqrt{2}}(-|001\rangle+|110\rangle)$ & Yes & 101 & 0.897 & 92.53\tabularnewline
\cline{4-9} \cline{5-9} \cline{6-9} \cline{7-9} \cline{8-9} \cline{9-9} 
 &  &  & $1^{st}$ \& $4^{th}$ & $\frac{1}{\sqrt{2}}(-|000\rangle+|111\rangle)$ & Yes & 100 & 0.955 & 95.37\tabularnewline
\cline{4-9} \cline{5-9} \cline{6-9} \cline{7-9} \cline{8-9} \cline{9-9} 
 &  &  & $2^{nd}$ \& $3^{rd}$ & $\frac{1}{\sqrt{2}}(-|000\rangle+|111\rangle)$ & Yes & 100 & 0.954 & 94.82\tabularnewline
\cline{4-9} \cline{5-9} \cline{6-9} \cline{7-9} \cline{8-9} \cline{9-9} 
 &  &  & $2^{nd}$ \& $4^{th}$ & $\frac{1}{\sqrt{2}}(-|001\rangle+|110\rangle)$ & Yes & 101 & 0.889 & 92.45\tabularnewline
\cline{4-9} \cline{5-9} \cline{6-9} \cline{7-9} \cline{8-9} \cline{9-9} 
 &  &  & $3^{rd}$ \& $4^{th}$ & $\frac{1}{\sqrt{2}}(-|001\rangle-|110\rangle)$ & Yes & 001 & 0.921 & 93.54\tabularnewline
\hline 
\multirow{4}{1cm}{\centering{4}} & $|\psi_{GHZ}\rangle$ & 3 & $1^{st}$, $2^{nd}$ \& $3^{rd}$ & $\frac{1}{\sqrt{2}}(-|010\rangle+|101\rangle)$ & Yes & 110 & 0.855 & 93.85\tabularnewline
\cline{4-9} \cline{5-9} \cline{6-9} \cline{7-9} \cline{8-9} \cline{9-9} 
 &  &  & $1^{st}$, $2^{nd}$ \& $4^{th}$ & $\frac{1}{\sqrt{2}}(-|011\rangle+|100\rangle)$ & Yes & 111 & 0.863 & {89.99}\tabularnewline
\cline{4-9} \cline{5-9} \cline{6-9} \cline{7-9} \cline{8-9} \cline{9-9} 
 &  &  & $1^{st}$, $3^{rd}$ \& $4^{th}$ & $\frac{1}{\sqrt{2}}(-|011\rangle-|100\rangle)$ & Yes & 011 & 0.889 & 93.41\tabularnewline
\cline{4-9} \cline{5-9} \cline{6-9} \cline{7-9} \cline{8-9} \cline{9-9} 
 &  &  & $2^{nd}$, $3^{rd}$ \& $4^{th}$ & $\frac{1}{\sqrt{2}}(-|010\rangle-|101\rangle)$ & Yes & 010 & 0.932 & 94.25\tabularnewline
\hline 
5 & $|\psi_{GHZ}\rangle$ & 4 & $1^{st}$, $2^{nd}$, $3^{rd}$ \& $4^{th}$ & $\frac{1}{\sqrt{2}}(-|000\rangle-|111\rangle)$ & No & 000 & 0.965 & 97.45\tabularnewline
\hline 
\end{tabular} 
\par\end{centering}
\caption{Theoretically expected results for the realization of quantum veto
using the GHZ state (Protocol B). Here, $|\psi_{GHZ}\rangle=\frac{1}{\sqrt{2}}(|000\rangle+|111\rangle)$. }.\label{tab:Result ghz}
\end{table}

\section{Comparison of Protocol A and Protocol B\label{sec:Comparison-of-Protocol}}

In the previous sections, we have shown that both Protocol A and Protocol
B can be experimentally realized using the IBMQ Casablanca with some
experimental errors. However, no analysis of the relative performance
of these protocols has yet been made. A comparative idea can be obtained
by comparing the range of fidelity as reported in the last columns
of Tables \ref{tab:Result Bell}, \ref{tab:Result cluster} and \ref{tab:Result ghz}.
We can easily observe that for Bell state based implementation of
Protocol A fidelity varies between 95.19 and 99.44, whereas the same
for cluster (GHZ) state based implementation of Protocol B is found
to vary between 88.79 and 97.63 (89.99 and 97.67). The obtained ranges
of fidelity clearly shows that Bell state based protocol performs better
than GHZ state based protocol and that in turn works better than cluster
state based protocol. This is not surprising as with the increase
in the number of qubits in the initial entangled state, difficulties
associated with the creation and maintenance of the entangled state
increases. This straight forward fidelity based analysis of the performance
of the protocols under ideal scenario can also be extended to a more
realistic situation involving noise. The same is done in the next
section.

\subsection{Effect of noise\label{sec:Effect-of-noise}}

\begin{figure}
\begin{centering}
\includegraphics[scale=0.6]{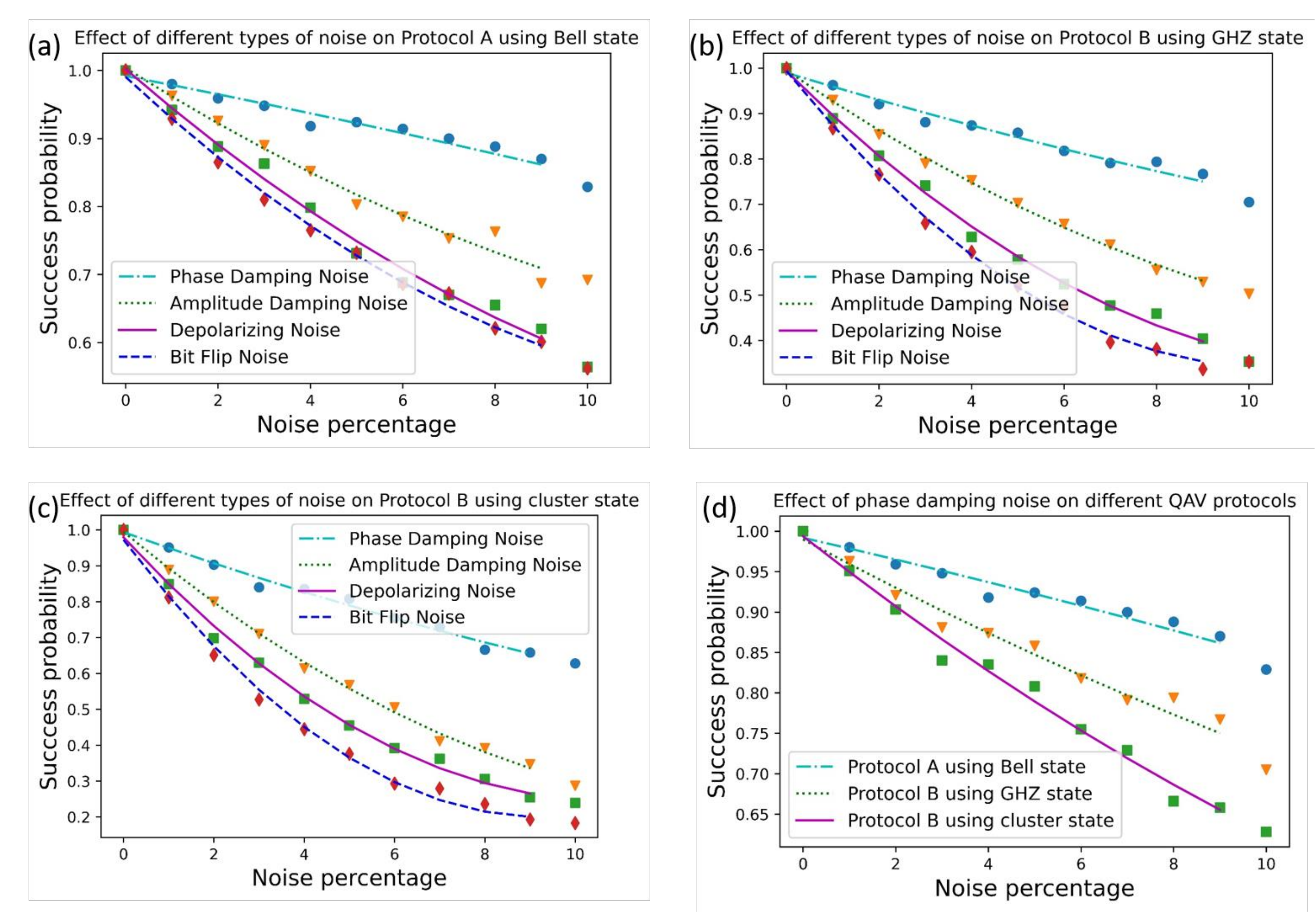}
\par\end{centering}
\centering{}\caption{Effect of phase damping, amplitude damping, depolarization and bit
filp noise on the (a) Protocol A using Bell state (b) Protocol B using
GHZ state (c) Protocol B using cluster state. (d) Effect of phase
damping noise on the Protocol A and the Protocol B. \label{fig:Effect-of-noise}}
\end{figure}

In realistic situations, quantum systems can never be perfectly isolated
from the surrounding environment and this leads to the degrading of
the unique quantum mechanical features. Hence, the practical implementations
of any quantum state based protocol necessarily involve the effects
of unwanted noise. So, any protocol is considered as practically useful
only if it is robust in the presence of noise (for details see \cite{banerjee2017asymmetricNoise,banerjee2018quantumNoise,sharma2016verificationNoise,thapliyal2018orthogonalNoise}).
Some of the most relevant noises that are usually considered in any
physical implementation are amplitude damping, phase damping, depolarizing
and bit flip noise. Further, the effect of the underlying noise on
the protocol can be studied in terms of the corresponding change in
the fidelity with respect to their noiseless implementations. The
effect of amplitude damping and phase damping noise on the quantum
anonymous veto protocols realized here has already been studied in
Ref. \cite{mishra2021quantumveto} from a different perspective. Here,
we have studied the effect of noise using noise model building technique
introduced on qiskit \cite{qiskit.noise}, an open source software
for working with quantum computer. Specifically, we have used the
error function available on qiskit to see the effect of different
types of noise. Effects of amplitude damping, phase damping, depolarizing
and bit flip noise on Protocol A and Protocol B (using both cluster
state and GHZ state) are illustrated in Fig. \ref{fig:Effect-of-noise}
(a)-(c). Each plot describes effect of 4 different types of noise on
a particular protocol. The plots are consistent with the earlier result
of Ref. \cite{mishra2021quantumveto} and clearly show that the effect
of different noise models can be arranged in an ascending order (based
on the diminishing effect of noise on Fidelity) as phase damping,
amplitude damping, depolarizing, bit-flip noise. Further, it's observed
that independent of the nature of noise, Bell state based schemes always
perform better than the other two schemes studied here. As an example,
effect of phase damping noise on Protocol A and two implementations
of Protocol B (using GHZ or cluster state) are illustrated in Fig.
\ref{fig:Effect-of-noise} (d). Similar characteristics are observed
for other noise models, too (not shown here).

\section{Conclusion\label{sec:Conclusion}}

The different types of voting strengthens democracy in different manners.
However, designing quantum protocols for voting is not easy. Still,
a set of protocols for quantum voting have been proposed and analyzed
since 2006. Recently, a special type of quantum voting called quantum
anonymous veto has drawn considerable attention of the community.
As the name suggests, such a scheme would allow a set of voters to
execute the right to veto in an anonymous manner. We have already
mentioned that such a situation where protocol for quantum anonymous
veto is useful exist in UN security council. Some of the present authors
have already designed a set of protocols for quantum anonymous veto.
Here, we have implemented two of those protocols (which are more efficient)
using a superconductivity based quantum computer named IBMQ Casablanca.
The protocols that we have selected for the present study are referred
to as Protocol A and Protocol B. In the above, we have reported Bell
state based experimental realization of Protocol A whereas Protocol
B has been experimentally realized using GHZ state and cluster state.
The protocols are realized with high success probability, and it's
understood that using the present technology, a protocol for quantum
anonymous veto can be realized experimentally if the number of people
who can veto remains small as in the case of UN security council.
Further, it's observed that Bell state based Protocol A implemented
here performs better than the GHZ/cluster state based implementation
of the Protocol B in ideal scenario as well as in the presence of
different types of noise (amplitude damping, phase damping, depolarizing
and bit-flip noise). Further, it's observed that based on diminishing
impact on fidelity, different noise models studied here can be ordered
in ascending order as phase damping, amplitude damping, depolarizing,
bit-flip.

\section*{Acknowledgment}

Authors acknowledge the support from the QUEST scheme of Interdisciplinary
Cyber Physical Systems (ICPS) program of the Department of Science
and Technology (DST), India (Grant No.: DST/ICPS/QuST/Theme-1/2019/14
(Q80)). They also thank Sandeep Mishra and Kishore Thapliyal for their
interest in the work.

\bibliographystyle{unsrt}
\bibliography{QV_reference}

\end{document}